\documentclass[11pt]{extarticle}
\usepackage{fullpage}
\usepackage[english]{babel}
\usepackage[utf8]{inputenc}
\usepackage{amsmath, amssymb, bm}
\usepackage{graphicx}
\usepackage{subcaption}
\usepackage[colorinlistoftodos]{todonotes}
\usepackage{indentfirst}

\newcommand{\hiddenpower}[2] { \ifnum \numexpr#2=1 #1 \else #1^#2 \fi }
\numberwithin{equation}{section}

\newcommand{\pd}{\partial}

\usepackage{xparse}
\ExplSyntaxOn
\newcounter{diff_order}
\newcounter{diff_power}

\newcommand{\rawdiff}[3]
{
	\setcounter{diff_order}{0}
	\clist_map_inline:nn{#3}{\stepcounter{diff_order}}
	
	\frac{\hiddenpower{#1}{\thediff_order} #2}
	{
		\def\old_var{DefaultValue}
		\setcounter{diff_power}{0}
		
		\clist_map_inline:nn{#3}
		{
			\def\new_var{##1}
			\ifnum \thediff_power=0
				\stepcounter{diff_power}
			\else
				\tl_if_eq:NNTF \new_var \old_var
				{\stepcounter{diff_power}}
				{
					#1 \hiddenpower{\old_var}{\thediff_power}
					\setcounter{diff_power}{1}
				}
			\fi

			\def\old_var{##1}
		}
		
		#1 \hiddenpower{\old_var}{\thediff_power}
	}
}

\newlength{\bibitemsep}
\newlength{\bibparskip}
\let\oldthebibliography\thebibliography
\renewcommand\thebibliography[1]{%
  \oldthebibliography{#1}%
  \setlength{\parskip}{\bibitemsep}%
  \setlength{\itemsep}{\bibparskip}%
}

\def\hybrid{\topmargin 0pt    \oddsidemargin 0pt 
        \headheight 0pt \headsep 0pt
        \textwidth 16.5cm      
        \textheight 22.2cm       
        \marginparwidth .875in
        \parskip 5pt plus 1pt   \jot = 1.5ex}

\hybrid

\newcommand{\pdiff}[2]{\rawdiff{\pd}{#1}{#2}}

\ExplSyntaxOff

\newcommand{\lb}{\left(}
\newcommand{\rb}{\right)}

\renewcommand{\sinh}[2][1]{\hiddenpower{\text{sinh}}{#1} \lb #2 \rb}

\renewcommand{\ln}[1]{\text{ln} \lb #1 \rb}
\newcommand{\e}[1]{\text{e}^{#1}}

\newcommand{\tr}[1]{\text{tr}\left\lbrace #1 \right\rbrace}
\newcommand{\ptr}[2]{\text{tr}_{#1}\left\lbrace #2 \right\rbrace}

\begin{document}


\begin{titlepage}
\begin{center}
\strut\hfill


\vskip 0.45in

\noindent {{\bf{THE QUANTUM AUXILIARY LINEAR PROBLEM\\ $\&$ DARBOUX-B\"ACKLUND TRANSFORMATIONS }}}\\
\vskip 0.4in
\noindent {{\footnotesize{ANASTASIA DOIKOU AND IAIN FINDLAY{\footnote{\tt E-mail: a.doikou@hw.ac.uk, iaf1@hw.ac.uk }}}}}
\vskip 0.04in
\noindent {\footnotesize Department of Mathematics, Heriot-Watt University,\\
Edinburgh EH14 4AS, United Kingdom}


\end{center}

\vskip .05in
\begin{abstract}
\noindent { We explore the notion of the quantum auxiliary linear problem and the associated problem of quantum B\"acklund transformations (BT). 
In this context we systematically construct the analogue of the classical formula that provides the whole hierarchy of the time components of 
Lax pairs at the quantum level for both closed and open integrable lattice models. The generic time evolution operator formula is particularly 
interesting and novel at the quantum level when dealing with systems with open boundary conditions. In the same frame we show that the reflection 
$K$-matrix can also be viewed as a particular type of BT, fixed at the boundaries of the system. The $q$-oscillator ($q$-boson) model, a variant 
of the Ablowitz-Ladik model, is then employed as a paradigm to illustrate the method. Particular emphasis is given to the time part of the quantum 
BT as possible connections and applications to the problem of quantum quenches as well as the time evolution of local quantum impurities are evident.
A discussion on the use of Bethe states as well as coherent states and the path integral formulation for the study of the time evolution is also presented.}
\end{abstract}
\date{}
\vskip 0.1in

\end{titlepage}


\section{Introduction}

\noindent One of the main purposes of the present investigation is the the study of the time evolution problem for discrete quantum integrable systems, 
i.e. the derivation of  the quantum analogue of the Semenov-Tian-Shansky formula \cite{sts} for both periodic and open boundary conditions \cite{avan-doikou1}. 
More precisely, the quantum hierarchy of the time components of the Lax pairs is extracted via the underlying quantum algebra. Based on the frame of the quantum 
auxiliary problem we then introduce the notion of quantum Darboux-B\"acklund transformations (BT). This is the first time, to our knowledge, that the issue of 
quantum Darboux-B\"acklund transformations is treated in the context of continuum time. To date, quantum BTs have been derived using the $Q$-operator setting \cite{sklyanin, korff} 
and are basically associated to integrable quantum systems with discrete time. Here we present the general setting and employ the system of $N$ $q$-oscillators as a paradigm to 
illustrate our formulation. It is worth noting that previous similar findings on the time independent part of the quantum BT \cite{korff}, obtained via the $Q$-operator approach 
for an analogous model, the quantum Ablowitz-Ladik lattice, are essentially recovered.

The proposed setting is closely related to the theme of quantum quenches \cite{caux}, given that the time evolution of the quantum observables for $N$-body systems is 
the question at hand (see also relevant recent results at the classical level \cite{CD}). Especially relevant in this context is the information one obtains from the time part 
of the quantum BT. Also, in view of recent results on the relation of the time part of the BT with the time evolution of local integrable defects \cite{corrigan-BT, avan-doikoub, doikou-BT}, 
it is clear that the time evolution is a particularly relevant issue in this setting as well. In this spirit the space-time duality established in \cite{ACDK} can be further explored,
specifically at the quantum level and in relation to systems with discrete space and time. Another associated problem of significance is the derivation of the quantum
 Gelfand-Levitan-Marchenko (GLM) equation (see e.g. \cite{FT, ablo-cla}). The GLM equation arises naturally via the Zakharov-Shabat dressing formulation 
\cite{ZS, Drazin} as part of a Darboux-type transformation \cite{Darboux}, where the involved quantities are now integral operators.

The outline of this article is as follows: In section 2 we introduce the concept of the quantum auxiliary linear problem for semi-discrete integrable 
systems with periodic boundary conditions. Employing the notions of quantum $R$-matrix and the underlying quantum algebra we rigorously 
derive the universal expression that provides the whole hierarchy of the quantum time components of the Lax pairs for the various time flows. 
Note that a similar formula is presented in \cite{korepin} for closed spin chains. Then, the periodic $q$-oscillator spin chain is considered as a paradigm, 
and the  quantum time components of the Lax pairs associated to first integrals of motion are explicitly constructed.
In section 3 we extend our analysis to the case where integrable boundary conditions are also incorporated. The universal expression 
for the time components of the Lax pairs is also constructed, and it turns out to have a distinctly different form compared to the classical 
analogue derived in \cite{avan-doikou1}. The corresponding open $q$-oscillator chain is then considered and explicit computations
 of boundary time components of the Lax pairs are performed. In section 4 the quantum B\"acklund transformation is discussed. 
We start our analysis from a generic Darboux matrix satisfying a certain algebraic structure and we then explicitly derive the quantum 
BT relations. We find both the time independent part, which is similar to the corresponding result in \cite{korff}, 
as well as the time dependent part of the BT relations. The time dependent expressions are particularly relevant, 
especially regarding the problem of integrable discontinuities on the real line. A brief discussion on the suitable quantum state picture,
 i.e. Bethe ansatz methodology as well as coherent state path integral formulation associated to the problem at hand is presented in section 5,
 together with a discussion on the main findings of this article as well as on possible future directions is given.

\section{Time evolution: the closed quantum spin chain}

\noindent Before we proceed to our main aim, which is the derivation of the time components of quantum  Lax pairs $(L_n,\ {\mathbb A}_n )$ 
let us recall the semi-discrete auxiliary linear problem \cite{FT}
\begin{equation}
\begin{aligned}
	\Psi_{n+1}(\lambda) &= L_n(\lambda)\ \Psi_n(\lambda), \\
	\partial_t \Psi_n(\lambda) &= {\mathbb A}_n(\lambda)\ \Psi_n(\lambda).
\end{aligned} \label{auxiliary}
\end{equation}
\noindent A similar discussion is provided in \cite{korepin} and the final expression formally resembles our findings. 
Here we are mostly interested in models with open boundary conditions, which will be discussed in detail in the subsequent section.

The fundamental structure underlying quantum integrable models comes from the Yang-Baxter equation \cite{korepin}
\begin{equation}
R_{ab}(\lambda - \mu) R_{ac}(\lambda) R_{bc}(\mu) = R_{bc}(\mu) R_{ac}(\lambda) R_{ab}(\lambda - \mu), \label{YBE}
\end{equation}
\noindent where $ R $ is a matrix that acts on two copies of an underlying vector space, $ V \otimes V $, with the subscripts 
denoting which two copies it acts on, so that the whole equation acts on $ V \otimes V \otimes V $. The $ R $-matrices are 
allowed to depend on some additional free parameter, denoted by either $ \lambda $ or $ \mu $, called the spectral parameter.

In the quantum setting (in the Heisenberg picture), the time evolution of the operators in a system is given by Heisenberg's equation:
$\pd_t \mathcal{O} =  \big [H,\  \mathcal{O}\big ]$,
where $ \mathcal{O} $ is the operator in question and $ H $ is the Hamiltonian. 
Thus, if we want to consider the time evolution of a system, we need to first find expressions for 
the commutators between the fields of the model. We do this through the RLL relation
\begin{equation}
R_{ab}(\lambda - \mu) L_{an}(\lambda) L_{bn}(\mu) = L_{bn}(\mu) L_{an}(\lambda) R_{ab}(\lambda - \mu). \label{RLL}
\end{equation}
 It will be beneficial to rewrite this in a more explicit manner
\begin{equation}
\Big[ L_{an}(\lambda),\  L_{bm}(\mu) \Big] = \Big( \mathfrak{R}_{ab}(\lambda - \mu) L_{an}(\lambda) L_{bn}(\mu) - 
L_{bn}(\mu) L_{an}(\lambda) \mathfrak{R}_{ab}(\lambda - \mu) \Big) \delta_{nm}, \label{eq:qAlg}
\end{equation}
\noindent where we define for compactness $ \mathfrak{R}_{ab}(\lambda) = \mathbb{I}_{ab} - R_{ab}(\lambda) $.
 The Kronecker delta factor is introduced to more generally describe the commutativity of matrices acting on wholly different spaces.

The generation of commuting quantities now follows , where we first define
\begin{equation}
T_a(n, m; \lambda) = L_{an}(\lambda) \cdots L_{am}(\lambda),
\end{equation}
\noindent with $ n > m $, and then the monodromy matrix as $ T_a = T_a(N, 1) $. 
The monodromy matrix also satisfies an fundamental relation (\ref{RLL}).
The transfer matrix $ \mathfrak{t} $ is then defined as the trace of the monodromy matrix, $ \mathfrak{t}(\lambda) = \ptr{a}{T_a(\lambda)} $, 
and by making use of the RTT relation we can see that this commutes with itself for different values of the spectral parameter. 
Consequently, if this is expanded as a power series in the spectral parameter, the coefficients $ \mathfrak{t}^{(k)} $ of $ \lambda^k $ will commute with one another:
$\big[ \mathfrak{t}^{(k)},\ \mathfrak{t}^{(j)} \big] = 0.$

As these coefficients all commute with one another, the logarithm of the transfer matrix can be considered, 
$ \mathcal{G}(\lambda) = \ln{\mathfrak{t}(\lambda)} $, and the coefficients in the series expansion of that commute with one another as well:
$\big[ \mathcal{G}^{(k)},\ \mathcal{G}^{(j)} \big] = 0.$
Each of the quantities generated in either of these ways can be treated as the Hamiltonian governing the evolution along a distinct time flow. 
Then, as we know that they commute with one another, this tells us that the other quantities will all be constant in this system.

In parallel to how the classical STS formula is derived, we start by using the transfer matrix in place of the Hamiltonian
\begin{equation}
\pd_{t} L_{bn}(\mu) =  \Big[ \mathfrak{t}(\lambda),\  L_{bn}(\mu) \Big],
\end{equation}
\noindent where $ t $ denotes the ``universal time" that contains all of the distinct time flows. Then, as the $ L_{an} $ 
commute with $ L_{bm} $ when both $ a \neq b $ and $ n \neq m $ 
(that is, they act ultra-locally), the only term in the $ \mathfrak{t} $ that interacts with the commutator will be $ L_{an}$
\begin{equation}
\pd_{t} L_{bn}(\mu) =  \mbox{tr}_{a}\Big \{T_a(N, n + 1; \lambda) \big[ L_{an}(\lambda),\ L_{bn}(\mu) \big] T_a(n - 1, 1; \lambda)\Big\}.
\end{equation}
\noindent Using the alternate form of the RLL relation \eqref{eq:qAlg}, we can evaluate this commutator
\begin{equation}
\begin{aligned}
\pd_{t} L_{bn}(\mu) &=  \mbox{tr}_{a}\Big \{T_a(N, n + 1; \lambda) \mathfrak{R}_{ab}(\lambda - \mu) T_a(n, 1; \lambda)\Big \} L_{bn}(\mu) \\
&\qquad- L_{bn}(\mu) \mbox{tr}_a\Big \{T_a(N, n; \lambda) \mathfrak{R}_{ab}(\lambda - \mu) T_a(n - 1, 1; \lambda)\Big \}, \label{gener}
\end{aligned}
\end{equation}
\noindent however, by comparing this with the compatibility condition of the two halves of the quantum auxiliary 
linear problem we have
\begin{equation}
\pd_t L_{an} = {\mathbb A}_{a n + 1} L_{an} - L_{an} {\mathbb A}_{an}. \label{eq:qZCC}
\end{equation}
\noindent It is thus evident that  the $ {\mathbb A} $-matrix is expressed as a
\begin{equation}
\mathbb{A}_{bn}(\lambda, \mu) = \mbox{tr}_{a}\Big \{T_a(N, n; \lambda) \mathfrak{R}_{ab}(\lambda - \mu) T_a(n - 1, 1; \lambda)\Big\},
\end{equation}
or by recalling the definition of the $ \mathfrak{R} $-matrix
\begin{equation}
\mathbb{A}_{bn}(\lambda, \mu) = \mathfrak{t}(\lambda) \mathbb{I} - \mathbb{B}_{bn}(\lambda, \mu), \label{eq:q_STS}
\end{equation}
where we define
\begin{equation}
\mathbb{B}_{bn}(\lambda, \mu) = \mbox{tr}_{a}\Big \{T_a(N, n; \lambda) R_{ab}(\lambda - \mu) T_a(n - 1, 1; \lambda)\Big \}. \label{eq:q_BGen}
\end{equation}

To find the $ {\mathbb A} $-matrix associated to the Hamiltonian $ \mathfrak{t}^{(k)} $, we simply expand \eqref{eq:q_BGen} about powers of 
$ \lambda $ and take the coefficient of order $ \lambda^k $, labelled $ \mathbb{B}_{bn}^{(k)} $, and combine it with the Hamiltonian as
\begin{equation}
\mathbb{A}_{bn}^{(k)}(\mu) = \mathfrak{t}^{(k)} \mathbb{I} -  \mathbb{B}_{bn}^{(k)}(\mu).
\end{equation}
\noindent If we now insert this into the zero-curvature condition, \eqref{eq:qZCC}, then the terms explicitly containing $ \mathfrak{t}^{(k)} $ 
are simply the right-hand side of Heisenberg's equation. Consequently, it follows that the $ \mathbb{B}_{bn}^{(k)} $ are only shifted by the action 
of the $ L $-matrix and leave $ L_{an} $ unchanged, i.e. $ L_{an} \mathbb{B}_{an}^{(k)} = \mathbb{B}_{a, n + 1}^{(k)} L_{an} $. This can in fact be 
shown to hold for the entire generator $ \mathbb{B}_{bn} $ by using the RLL relation \eqref{RLL}
\begin{equation}
L_{an}(\mu) \mathbb{B}_{an}(\lambda, \mu) = \mathbb{B}_{a n + 1}(\lambda, \mu) L_{an}(\mu). \label{eq:BLComm}
\end{equation}
As mentioned when defining the generator of the commuting quantities, we are actually interested in the quantities generated by $ \mathcal{G}(\lambda) $. 
Unfortunately, however, the process for finding the $ {\mathbb A} $-matrix generator in this quantum setting is not as simple as it was for the equivalent classical case, 
since we can no longer use the fact that $ \big [\ln{a},\ b\big ] = a^{-1} \big [a,\ b\big ] $ due to the non-commutativity of $ \mathfrak{t} $ 
and $ [\mathfrak{t},\ \mathcal{O}] $.
Nevertheless, suitable combinations of the non-local ${\mathbb A}$-operators, in analogy to the classical case, can provide the local ones, 
compatible also with the equations of motion
coming form the corresponding local quantum Hamiltonians.

\subsection{Application: the $q$-harmonic oscillator}

\noindent To illustrate the setting described in the previous subsection in practice, we choose to consider as an example the $q$-harmonic oscillator, 
which provides a variation of the quantum Ablowitz-Ladik model, as well as a lattice version of the quantum NLS model, and is also related to the Liouville model. 
The associated Lax operator is given by
\begin{equation}
	L_n(\lambda) = \lb \begin{matrix}
		u v_n & a^{\dag}_n \\
		a_n & -u^{-1} v_n
	\end{matrix} \rb, \label{eq:AL_L}
\end{equation}
\noindent where $ u = \e{\lambda} $. It is convenient in what follows to introduce the fields $ b_n = v_n^{-1} a_n $ and $ b^{\dag}_n = v_n^{-1} a^{\dag}_n $. 
We then use this in the RLL relation, with the familiar XXZ $ R $-matrix \cite{jimbo}
\begin{equation}
\begin{gathered}
	R(\lambda) = \alpha \sum_{i=1}^2e_{ii}\otimes e_{ii} + \beta \sum_{i\neq j=1}^2 e_{ii}\otimes e_{jj}  + \gamma \sum_{i\neq j=1}^2 e_{ij}\otimes e_{ji}, \nonumber\\
	\alpha = q u - q^{-1} u^{-1}, ~~~~\beta =u-u^{-1}, ~~~~\gamma = q-q^{-1},
\end{gathered} \label{eq:AL_R}
\end{equation}
\noindent where $q= e^{\hat \mu}$, and we define the generic $N \times N$ matrix $e_{ij}$ ($2\times 2$ in our case) with elements $(e_{ij})_{kl} = \delta_{ik}\ \delta_{jl}$. 
Hence, we obtain the following commutation relations
\begin{equation}
\begin{aligned}
	\Big [ b_n, b^{\dag}_m \Big] &= (q - q^{-1}) v_n^{-2} \delta_{nm}, \\
	\Big [ b_n, v_m \Big ] &= (1 - q) b_n v_n \delta_{nm}, \\
	\Big [ b^{\dag}_n, v_m \Big ] &= (1 - q^{-1}) b^{\dag}_n v_n \delta_{nm}, \\
	\Big [ b_n, b_m \Big ] &= \left [ b^{\dag}_n, b^{\dag}_m \right] = \Big [ v_n, v_m \Big] = 0.
\end{aligned} \label{eqs:AL_Comm}
\end{equation}
\noindent Indeed, by expanding the trace of the monodromy matrix about powers of $ u $, we can find the Hamiltonians for the system. 
Due to the symmetry in the $ L_n $ matrix, we have a choice of sending $ \lambda $ to either plus or minus infinity, corresponding to 
the limits $ u \to \infty $ and $ u \to 0 $ respectively. In each of these cases, we will get a slightly different tower of 
Hamiltonians (labelled $ H^{(+, k)} $ and $ H^{(-, k)} $ respectively), and the physical Hamiltonian can be seen to 
be constructed from the sum $ H =q H^{+} +q^{-1} H^{-} $, where $ H^{+} = (H^{(+, 0)})^{-1} H^{(+, 2)} $ and $ H^{-} = (H^{(-, 0)})^{-1} H^{(-, 2)} $.
 Evaluating this, we get that the Hamiltonians $H^{\pm}$ are
\begin{equation}
	H^+ = \sum_{j = 1}^{N} b^{\dag}_{n + 1} b_n, ~~~~ H^- =  \sum_{j = 1}^{N}b_{n + 1} b^{\dag}_n. \label{eq:AL_H}
\end{equation}
It is clear that any linear combination of $H^{\pm}$ will also provide an integral of motion.
We can now derive the associated ${\mathbb A}$-operator (details on the computations, and in particular the expressions for 
the ${\mathbb B}$-operator are provided in Appendix A). The associated $ {\mathbb A}^{\pm}_n $ matrices read as
\begin{equation}
	{\mathbb A}^+_n = \lb \begin{matrix}
		\zeta u^2 +{\cal A} b_n^{\dag}b_{n-1}  & u {\cal B} b^{\dag}_n\\
		u {\cal C} b_{n-1} & {\cal D} b_n^{\dag} b_{n-1}
	\end{matrix} \rb,  ~~~~~{\mathbb A}^-_n = \lb \begin{matrix}
		 \tilde {\cal A} b_{n-1}^{\dag}b_{n}  & u^{-1} \tilde  {\cal B} b^{\dag}_{n-1}\\
		u^{-1} \tilde {\cal C} b_{n} & \tilde \zeta u^{-2} + {\cal D} b_n^{\dag} b_{n-1}
	\end{matrix} \rb, \label{eq:DarbouxAA	}
\end{equation}
\noindent where:
\begin{equation}
\begin{gathered}
	\zeta = {\cal B} = q^{-2} -1, ~~~\tilde \zeta = \tilde {\cal C} =q^2-1,~~~{\cal A} = \tilde {\cal D} =1-q^{-1}, \\
	{\cal C} = \tilde {\cal B}=  q^{-1} - q,~~~{\cal D} = \tilde {\cal A} = 1 - q.
\end{gathered}
\end{equation}

Now that we have both the Hamiltonian \eqref{eq:AL_H}, and the complete Lax pair, we can find the time evolution of the 
fields $ v_n $, $ b_n $, and $ b^{\dag}_n $ associated to the sum $H =qH^+ +q^{-1}H^-$. This choice is convenient as 
will become transparent when studying the open spin chain in the subsequent section. Only one of the two 
approaches is necessary (either through Hamilton's equations $ \dot{L}_n = \big [H,\  L_n\big ] $ 
or the zero curvature condition $ \dot{L}_n = {\mathbb A}_{n + 1} L_n - L_n  {\mathbb A}_n $) as they both yield the same time evolution, namely
\begin{equation}
\begin{aligned}
	\dot{v}_n &= (1-q)  v_n b^{\dag}_n (q^{-1}b_{n + 1} + q b_{n - 1}) - (1 - q) v_n b_n (qb^{\dag}_{n + 1} + q^{-1}b^{\dag}_{n - 1}), \\
	\dot{b}_n &= (q^{-1} - q) v_n^{-2} (q^{-1}b_{n + 1} + qb_{n - 1}), \\
	\dot{b}^{\dag}_n &= (q - q^{-1}) v_n^{-2} (qb^{\dag}_{n + 1} +q^{-1} b^{\dag}_{n - 1}).
\end{aligned} \label{eqs:AL_EoMs}
\end{equation}

\section{Time evolution: the open quantum spin chain}

\noindent We are particularly interested in the case when integrable boundary conditions are also incorporated. 
The expressions for the hierarchy of the time components of the Lax pairs is novel at the quantum level, 
and has a non-trivial form compared to the classical case derived in \cite{avan-doikou1}. Similar reasoning 
can be applied to the open spin chain, where now one takes also into account the left 
and right reflection matrices, $ K^{+} $ and $ K^{-} $,  where $ K^{-} $ satisfies the reflection algebra \cite{cherednik, sklyaninb}
\begin{equation}
	R_{12}(\lambda - \mu) K^{-}_1(\lambda) R_{21}(\lambda + \mu) K^{-}_2(\mu) = K^{-}_2(\mu) 
R_{12}(\lambda + \mu) K^{-}_1(\lambda) R_{21}(\lambda - \mu), \label{eq:RE}
\end{equation}
\noindent and $K^{+}(\lambda) = M \big( K^{-}(-\lambda - \rho \big)^T$ for some matrix $ M $ that satisfies 
$\big  [R_{12},\ M_1 M_2\big ] = 0 $. In our example in the next subsection we are going to focus on the two dimensional case, 
$M={\mathbb I}$ and $\rho = \hat \mu$, 
recall also that $q = e^{\hat \mu}$.

With these extra matrices, a modified monodromy matrix is derived \cite{sklyaninb}, which also satisfies the reflection algebra above
\begin{equation}
	\mathcal{T}(\lambda) = T(\lambda) K^{-}(\lambda) \hat{T}(-\lambda) K^{+}(\lambda), \label{eq:RE_T}
\end{equation}
\noindent where we define $\hat{T}_0(\lambda) = V_0\ T_0^{t_0}(-\lambda -\rho)\ V_0$, where
$V$ is a constant matrix suh that $V^2 ={\mathbb I}$. In the next subsection we will focus on the two dimensional 
case, where $\rho = \hat \mu$ and $ V = \mbox{antidiag} (1,1) $.

Indeed, we shall use the equivalent of \eqref{RLL} for the $ T $ and $ \hat{T} $ matrices. As
will introduce factors of ${\mathfrak R}_{ab}(\lambda - \mu) $ into the monodromy matrices, we will introduce the notation
 that $ T^{+}_a = T_a(N, n + 1; \lambda) $ and $ T^{-}_a = T_a(n - 1, 1; \lambda) $.  With these, we can evaluate the commutator 
of $ \mathfrak{t} $ with $ L_{bn} $ to get the time evolution of $ L_{bn} $, where for brevity, we shall refer to 
$ {\mathfrak R}_{ab}(\lambda - \mu) $ as $ {\mathfrak R}^{-}_{ab} $ and $ {\mathfrak R}
_{ab}(\lambda + \mu) $ as $ {\mathfrak R}^{+}_{ab} $
\begin{align}
	\dot{L}_{bn} &= \ptr{a}{T^{+}_a {\mathfrak R}_{ab}^{-}  L_{an} T^{-}_a L_{bn} K^{-}_{a} \hat{T}_a K^{+}_{a}} - 
\ptr{a}{L_{bn} T^{+}_a L_{an} {\mathfrak R}_{ab}^{-} T^{-}_a K^{-}_{a} \hat{T}_a K^{+}_{a}} \nonumber \\
	& \qquad + \ptr{a}{T_a K^{-}_{a} \hat{T}^{-}_a \hat{L}_{an} {\mathfrak R}_{ab}^{+}  \hat{T}^{+}_a L_{bn} K^{+}_{a}} - 
\ptr{a}{T_a K^{-}_{a} L_{bn} \hat{T}^{-}_a {\mathfrak R}_{ab}^{+} \hat{L}_{an} \hat{T}^{+}_a K^{+}_{a}}. \nonumber
\end{align}
\noindent In order to compare the latter expression with the discrete zero curvature condition, we need to commute the $ L_{bn} $
 in the first and fourth through the $ \hat{T}_a $ and $ T_a $ respectively. Using the suitable commutators we find that the terms with the $ L_{bn} $
 still in between the monodromy matrices cancel out, leaving
\begin{align}
	\dot{L}_{bn} &= \ptr{a}{T^{+}_a {\mathfrak R}^{-}_{ab}  L_{an} T^{-}_a K^{-}_{a} \hat{T}_a K^{+}_{a}} L_{bn} - L_{bn}
\ptr{a}{T^{+}_a L_{an} {\mathfrak R}^{-}_{ab}  T^{-}_a K^{-}_{a} \hat{T}_a K^{+}_{a}} \nonumber \\
	&\qquad - L_{bn} \ptr{a}{T_a K^{-}_{a} \hat{T}^{-}_a {\mathfrak R}^{+}_{ab}  \hat{L}_{an} \hat{T}^{+}_a K^{+}_{a}} + 
\ptr{a}{T_a K^{-}_{a} \hat{T}^{-}_a \hat{L}_{an} {\mathfrak R}^{+}_{ab} \hat{T}^{+}_a K^{+}_{a}} L_{bn} \nonumber \\
	&\qquad- \ptr{a}{T^{+}_a {\mathfrak R}^{-}_{ab}  L_{an} T^{-}_a K^{-}_{a} \hat{T}^{-}_a \hat{L}_{an} {\mathfrak R}^{+}_{ab}
\hat{T}^{+}_a K^{+}_{a}}L_{bn} \nonumber \\
	&\qquad+ L_{bn} \ptr{a}{T^{+}_a L_{an} {\mathfrak R}^{-}_{ab} T^{-}_a K^{-}_{a} \hat{T}^{-}_a {\mathfrak R}^{+}_{ab}  
\hat{L}_{an} \hat{T}^{+}_a K^{+}_{a}}. \nonumber
\end{align}
\noindent If we compare this to the discrete zero curvature condition ($ \dot{L}_n =  {\mathbb A}_{n + 1} L_n - L_n  {\mathbb A}_n $), we can read off the 
expression for the ${\mathbb A}$-operator, and if we split the $ \mathbb{I} - R_{ab} $ terms, this simplifies to
\begin{equation}
	\mathbb{A}_n = \ptr{a}{T_a K^{-}_{a} \hat{T}_a K^{+}_{a}} - \ptr{a}{T^{+}_a L_{an} R^{-}_{ab} T^{-}_a K^{-}_{a} 
\hat{T}^{-}_a R^{+}_{ab} \hat{L}_{an} \hat{T}^{+}_a K^{+}_{a}}. \label{eq:qA_Open}
\end{equation}

Similarly to the case of the periodic chain, this consists of two terms, the first of which is just the generator of the Hamiltonians. 
Again, this means that each of the individual $ {\mathbb A}_n $ matrices can be written as the combination 
$ {\mathbb A}^{(k)}_n = H^{(k)}{\mathbb I} - {\mathbb B}^{(k)}_n $. 
Notice however the non-trivial structure of the second term of (\ref{eq:qA_Open}), which is quadratic in $R$ 
as opposed to the periodic case studied earlier in the text. This is not surprising given that it reflects the structure
of the underlying quantum algebra provided by the reflection equation, which is also quadratic in $R$. In the periodic 
case both classical and quantum expressions have the same structure due to the linearity of the underlying algebras in $R$. 
Although the open quantum case is distinctly different to the classical one, the classical limit of (\ref{eq:qA_Open}) 
naturally leads to the linear expression derived in \cite{avan-doikou1}.

In studying the open spin chain, we have to work with both the $ L_n $ matrices and $ \hat{L}_n $ matrices. It is clear the $ \hat{L}_n $
 are part of their own Lax pair $ (\hat{L}_n,\ \hat{\mathbb A}_n) $ satisfying the auxiliary linear problem
\begin{equation}
\begin{aligned}
	& \Psi_n = \hat{L}_n \Psi_{n + 1}, \\
	& \partial_t \Psi_n = \hat{\mathbb A}_n \Psi_n,
\end{aligned} \label{aux2}
\end{equation}
\noindent the compatibility condition of which gives the corresponding zero curvature condition. Consequently, we may be interested in
 finding the generator $ \hat{\mathbb{A}}_n $ for a closed chain of such $ \hat{L}_n $, much as we did for the normal $ L_n $. 
This is a particularly relevant issue as will become clear below when deriving the $ K $ matrices as fixed BTs at the boundaries of the system.

To derive the $ \hat {\mathbb A}_n $ we follow much the same procedure, except starting from the equation
\begin{equation}
	\dot {\hat L}_n = \Big [ \hat{\mathfrak{t}},\  \hat{L}_n \Big]. \nonumber
\end{equation}
\noindent After repeating all of the previous steps, we find that the generator is given by
\begin{equation}
	\hat{\mathbb{A}}_n(\lambda, \mu)  = \hat{\mathfrak{t}}(\lambda) - \ptr{a}{\hat{T}_a(1, n - 1; \lambda)
 \hat{R}_{ba}(\lambda - \mu) \hat{T}_a(n, N; \lambda)}, \label{eq:hat_qA}
\end{equation}
\noindent and we find that $ \hat{\mathbb{A}}_n = \mathbb{A}_n $, as the trace is invariant under both transposition and conjugation.

The transfer matrix $ \mathfrak{t} $ should naturally be constant with respect to time. In the open spin chain case, we can therefore 
use this to find relations between the reflection matrices $ K^{\pm} $ and the Lax pairs, $ (L_n, {\mathbb A}_n) $ and $ (\hat{L}_n, \hat{\mathbb A}_n) $. 
We take first the derivative of the transfer matrix, and also find from the zero curvature condition that $ \dot{T} = {\mathbb A}_{N + 1} T - 
T {\mathbb A}_1 $, and $ \dot{\hat{T}} = \hat{\mathbb A}_1 \hat{T} - \hat{T} \hat{\mathbb A}_{N + 1} $. Inserting these results into our 
time derivative of the transfer matrix, and after grouping the terms in a suggestive manner, we get that
\begin{equation}
	\dot{\mathfrak{t}} = \tr{(\dot{K}^{-} - {\mathbb A}_1 K^{-} + K^{-} \hat{\mathbb A}_1) \hat{T} K^{+} T + (\dot{K}^{+} + 
K^{+} {\mathbb A}_{N + 1} - \hat{\mathbb A}_{N + 1} K^{+}) T K^{-} \hat{T}}. \nonumber
\end{equation}
\noindent We shall consider the semi-infinite chain $N\to \infty$, so we are mostly interested in the boundary attached to the 
first site of the chain. Indeed,  the time derivative of the transfer matrix is zero, provided that
\begin{equation}
	\dot{K}^{-} = {\mathbb A}_1 K^{-} - K^{-} \hat{\mathbb A}_1.
\label{eq:K_BTs}
\end{equation}
\noindent The latter has the appearance of the time part of a BT, with $ K^{-} $ being a Darboux-type matrix. 
Note that in this case the quantities $ {\mathbb A}_n $ and $ \hat{\mathbb A}_n $ are somehow related via reflection, 
given the underlying algebraic construction of the modified monodromy matrix.

\subsection{The open $q$-oscillator model}

\noindent Once again, we shall use the open $q$-oscillator chain to test out this novel formulation. 
The $ L_n $ matrix and $ R $-matrix are the same as they were for the closed spin chain  \eqref{eq:AL_L} and 
\eqref{eq:AL_R} respectively, though we now also need to choose appropriate $ K^{\pm} $ matrices. 
We will only look at the simplest choice, $ K^{\pm} = \mathbb{I} $. It is worth pointing out that $ \hat{L}_n(\lambda)= L_n^{-1}(-\lambda) $, 
which can easily be shown by recalling the Casimir $ a^{\dag}_n a_n + q v_n^2 = 1 $.

Using the Casimir we can also see that $ H^{-} = q^2 H^{+} $ (see Appendix B), so the Hamiltonians in these two limits are equivalent. 
The benefit of this is that now when trying to find the corresponding $ {\mathbb A}_n $ matrix, we only need to look in one of the two
 limits (choosing $ H = qH^{+} = q^{-1} H^- $)
\begin{equation}
	H = \sum_{n = 1}^{N - 1} (qb^{\dag}_{n + 1} b_n + q^{-1} b_{n + 1} b^{\dag}_n) + 
(qb^{\dag}_1 b_1 + q^{-1} b_N b^{\dag}_N). \label{eq:AL_HOpen}
\end{equation}
\noindent As expected, this Hamiltonian is almost identical to the Hamiltonian \eqref{eq:AL_H} of the closed chain, up to boundary terms. 
Therefore the bulk $ {\mathbb A}_n $ matrices should also be the same; explicit computation confirms this. Let us present the $ {\mathbb A}_n $ 
matrix at the boundary i.e. $ {\mathbb A}_1 = H {\mathbb I}- {\mathbb B}_1 $ (see Appendix B for detailed computations).

Seeing as the Hamiltonian \eqref{eq:AL_HOpen} was found by considering $ qH^{+} $, the corresponding $ {\mathbb A}_n $ 
matrix will be found by considering $ {\mathbb A}_n = qH^{+}{\mathbb I} - q{\mathbb B}_n^{+} $. For the boundary case $ n = 1 $, the $ {\mathbb A}_1 $ matrix is then given by
\begin{equation}
	{\mathbb A}_1 = (u^2 + u^{-2}) \lb \begin{matrix}
		q^{-1} & 0 \\
		0 & q
	\end{matrix} \rb + \lb \begin{matrix}
		(q - q^{-1}) b^{\dag}_1 b_1 & (u + u^{-1})(q^{-1} - q) b^{\dag}_1 \\
		(u + u^{-1})(q^{-1} - q) b_1 & (q^{-1} - q) b_1 b^{\dag}_1
	\end{matrix} \rb. \label{eq:AL_qA1Open}
\end{equation}


As we did in the case of the closed spin chain, we can use the boundary $ {\mathbb A} $-operators to find the equations of motion at the boundaries. 
We shall look at the boundary, $ n = 1 $ (recall that we are considering here the semi-infinite chain $N \to \infty$).
 Then the equations of motion at the boundary read as
\begin{equation}
\begin{aligned}
	\dot{v}_1 &= (1 - q) v_1 b^{\dag}_2 b_1 + (1 - q^{-1}) v_1 b_2 b^{\dag}_1, \\
	\dot{b}_1 &= (q^{-2} - 1)(b_1 + b_2) v_1^{-2}, \\
	\dot{b}^{\dag}_1 &= (q^2 - 1)(b^{\dag}_1 + b^{\dag}_2) v_1^{-2}. \\
\end{aligned} \label{eqs:AL_EoM1Open}
\end{equation}
\noindent The results above agree with those found using the Heisenberg equation. Note also that the choice of 
$K^{-} \propto {\mathbb I}$ automatically satisfies the BT like relations for the $K$-matrix. 
A full classification of the $K$-matrices that satisfy (\ref{eq:RE}), and comparison with known (non) 
dynamical reflection matrices from the reflection equation is an appropriate issue, which however will 
be discussed in detail elsewhere (see also \cite{corrigan-BT} for a relevant discussion).

\section{Quantum B\"acklund Transformations}

\noindent We are now in a position to compute the quantum B\"acklund transformation for the $q$-oscillator model.
Recall first the Darboux transformation that connects two different auxiliary functions (see \cite{Darboux})
\begin{equation}
	\tilde \Psi_n = {\mathbb M}_n(\lambda)\ \Psi_n.
\end{equation}
\noindent Provided that the transformed auxiliary function $\tilde \Psi_n$ satisfies the auxiliary linear problem (with transformed
 $\tilde L_n,\ \tilde {\mathbb A}_n$), the following fundamental equations that lead to the associated B\"acklund transformation are obtained:
\begin{equation}
\begin{aligned}
	&  {\mathbb M}_{n+1}(\lambda)\ L_n(\lambda) = \tilde L_n(\lambda)\  {\mathbb M}_n(\lambda), \\
	& \pdiff{}{t}  {\mathbb M}_n(\lambda) = \tilde {\mathbb A}_n(\lambda)\  {\mathbb M}_n(\lambda) - {\mathbb M}_n(\lambda){\mathbb A}_n(\lambda).
\end{aligned} \label{eq:BT}
\end{equation}

Consider the following Darboux matrix (see also \cite{korff, doikou-findlay, suris})
\begin{equation}
	 {\mathbb M}_n = \lb \begin{matrix}
		\e{\lambda-\Theta} A_n - \e{-\lambda+\Theta} A_n^{-1} & X_n \\
		Y_n & -\e{-\lambda+\Theta}A_n
	\end{matrix} \rb. \label{eq:Darboux}
\end{equation}
\noindent It is important to note that in our construction here both $L_n$ and $\tilde{L}_n$ satisfy the same RLL algebra (\ref{RLL}). 
In fact, we choose to consider here the $q$-harmonic oscillator $L_n$ (\ref{eq:AL_L}), whereas $\tilde{L}_n$ is essentially the same operator, 
but with $a_n \to \tilde a_n,\ a^{\dag}_n \to \tilde a^{\dag}_n $. Notice that the Darboux matrix chosen above has essentially a similar algebraic structure as $L_n$. 
This is not particularly surprising given that the first of equations (\ref{eq:BT}) leads to the following formal expression for the Darboux matrix $ {\mathbb M}_n$
\begin{equation}
	 {\mathbb M}_{n+1}(\lambda, \{\Theta_i\}) = \tilde T(\lambda, \{\Theta_i\})\ {\cal M}(\lambda)\ T^{-1}(\lambda, \{\Theta_i\})
\end{equation}
\noindent where we define:
\begin{equation}
\begin{aligned}
	&\tilde T(\lambda) = \tilde L_n(\lambda, \Theta_n)\ \tilde L_{n-1}(\lambda, \Theta_{n-1}) \ldots \tilde L_{1}(\lambda, \Theta_{1}) \\
	&T^{-1}(\lambda) = L^{-1}_1(\lambda, \Theta_1)\  L^{-1}_{2}(\lambda, \Theta_{2}) \ldots L^{-1}_{n}(\lambda, \Theta_{n})
\end{aligned}
\end{equation}
\noindent and formally one can define ${\cal M}$ as a $c$-number matrix, ${\cal M} = \tilde L_0\ {\mathbb M}_0\ L^{-1}_0$. 
Of course the analytic structure of the $\tilde T_n$ and $T^{-1}_n$ matrices must be also taken into account when identifying 
the quantum BT (see also \cite{Darboux}). In order to explicitly identify the algebraic relations obeyed by the Darboux matrix $M_n$ 
a set of algebraic relations between $L_n$ and $\tilde{L}_n$ is required, as is the case for instance in reflection algebras \cite{cherednik, sklyaninb} 
and generic quadratic algebras \cite{maillet}, or in the context of integrable defects \cite{doikou-BT, avan-doikoub}.

Let us now consider the $t$-independent part of the BT relations. The basic relations arising from the time independent part of \eqref{eq:BT} 
are given by
\begin{equation}
\begin{aligned}
	&A_{n+1} v_n = \tilde v_n A_n, \\
	&X_n = \e{\Theta} A_n b_n^{\dag}, ~~~~~X_{n+1} = \e{-\Theta} (A_{n+1}^{-1}b^{\dag}_n - \tilde b^{\dag}_n A_{n+1}  ), \\
	&Y_{n+1} = \e{\Theta} \tilde b_n A_{n+1}, ~~~~~Y_n =  \e{-\Theta} (A_{n} b_n - \tilde b_n A_n^{-1}).
\end{aligned} \label{eq:BT*}
\end{equation}
\noindent Note also the Casimir operator (quantum determinant) associated to the Darboux matrix
$ M_n $
\begin{equation}
	q A_n^2 +X_n Y_n = q^{-1} A_n^2 + Y_n X_n =1, \nonumber
\end{equation}
\noindent which gives:
\begin{equation}
	A_n^{-2} = q  +\e{2\Theta} b^{\dag}_n \tilde b_{n-1}. \label{eq:casimir}
\end{equation}
\noindent Suitably comparing equations \eqref{eq:BT*} and taking into account \eqref{eq:casimir} we obtain the time independent part of the BT relations
\begin{equation}
\begin{gathered}
	q b_n^{\dag} -A_{n+1}^{-1} \tilde b_n^{\dag} A_{n+1}  = \e{2\Theta} b_{n+1}^{\dag} \Big (1- \tilde b_n b_{n}^{\dag} \Big ), \\
	q \tilde b_n  -q^{-1} A_n b_n A_n^{-1} = -\e{2\Theta}  \Big (1 + \tilde b_n b_n^{\dag} \Big) \tilde b_{n-1}.
\end{gathered} \label{eq:bti}
\end{equation}

The latter relations are similar to the ones found in \cite{korff} based on the $Q$-operator approach \cite{sklyanin}. 
From this point of view the system under study is a discrete time system, and the $Q$-operator is the generating function of the quantum BT \cite{sklyanin}. 
Our perspective here is rather different given that we are interested in continuum time systems, so time evolution in the Heisenberg picture is the problem at hand.  
Note also a technical observation; here the element $A_n$ of the Darboux matrix -- although expressed in terms of $ b_n^{\dag}, \tilde b_{n-1} $
-- is still apparent in the final expressions of the B\"acklund transformation as opposed to the case considered in \cite{korff}. 
This is essentially due to the fact that a different $R$-matrix is considered here, and the co-product structure of the underlying algebra is thus modified. 
In any case, the similarity between the expressions is apparent.

As already noted we are mostly interested in the continuum time picture of the problem. So in addition to the time independent relations \eqref{eq:bti} 
we shall derive below the time dependent part of the BT, in analogy to the classical case. To achieve this we focus on the second equation
 of the BT relations. We derived previously the time components of the Lax pairs for the $H^{\pm}$ Hamiltonians, 
thus below we shall derive two sets of time related equations for $ {\mathbb A}^{\pm}_n $ via the time part of the BT. 
In particular, the set of equations associated to $ {\mathbb A}^{+}_n $
\begin{equation}
\begin{aligned}
	&\dot X_n = {\cal A} \tilde b^{\dag}_n \tilde b_{n-1} X_n -{\cal D} X_n b_n^{\dag} b_{n-1} + {\cal B} \e{-\Theta} \Big ( A_n^{-1} b_n^{\dag} - \tilde b_n A_n\Big ), \\
	&\dot Y_n = {\cal D} \tilde  b^{\dag}_n \tilde b_{n-1} Y_n - {\cal A} Y_n b_n^{\dag} b_{n-1} + {\cal C} \e{-\Theta} \Big ( A_n b_{n-1} - \tilde b^{\dag}_{n-1} A_n^{-1} \Big ), \\
	&\dot A_n = {\cal D} \Big ( \tilde b^{\dag}_n \tilde b_{n-1} A_n - A_n b_n^{\dag} b_{n-1} \Big ),~~~-A_n^{-2} \dot A_n = {\cal A} \Big ( \tilde b^{\dag}_n \tilde b_{n-1} A_n^{-1} - A_n^{-1} b_n^{\dag} b_{n-1} \Big ).
\end{aligned} \label{eq:BTa}
\end{equation}
\noindent The second and third equations of the time independent part of the BT \eqref{eq:BT*} are also recovered.

Similarly, the relations associated to $ {\mathbb A}^{-}_n $ are given as
\begin{equation}
\begin{aligned}
	&\dot X_n = \tilde {\cal D} \tilde b_n \tilde b^{\dag}_{n-1} X_n -\tilde {\cal A} X_n b_n b^{\dag}_{n-1} -\tilde {\cal B} \e{\Theta} A_n b_{n-1}^{\dag}, \\
	&\dot Y_n = \tilde {\cal A} \tilde  b_n \tilde b^{\dag}_{n-1} Y_n - \tilde {\cal D} Y_n b_n b^{\dag}_{n-1} + {\cal C} \e{\Theta} \tilde b_{n} A_n, \\
	&\dot A_n = {\cal D} \Big ( \tilde b_n \tilde b^{\dag}_{n-1} A_n - A_n b_n b_{n-1}^{\dag} \Big ),~~~-A_n^{-2} \dot A_n = {\cal A} \Big ( \tilde b^{\dag}_n \tilde b_{n-1} A_n^{-1} - A_n^{-1} b_n^{\dag} b_{n-1} \Big ).
\end{aligned} \label{eq:BTb}
\end{equation}
\noindent The third and fourth equations in \eqref{eq:BT*} are now recovered. It is thus clear that via the time part of the BT for both $ {\mathbb A}^{\pm}_n $ 
all the time independent relations are reproduced, which suggests that in this particular study the time part provides all the required information.

The set of equations above give rise to more explicit time equations. For instance, focusing on \eqref{eq:bti} and \eqref{eq:BTa}, the following expressions are obtained
\begin{equation}
\begin{gathered}
	\dot{b}^{\dag}_n = (q-q^{-1}) A_n^{-1}\tilde b^{\dag}_n\tilde b_{n-1} A_n b^{\dag}_n + (q^{-2} -1) e~^{-2\Theta} (A_n^{-2}b^{\dag}_n - A_n^{-1}\tilde b_n A_n), \\
	\dot{b}_{n-1} = (q^{-1}-q)\tilde b_{n-1} A_n b^{\dag}_{n} b_{n-1} + (q^{-1} -q) e~^{-2\Theta} (A_n b_{n-1}A_n^{-1} - \tilde b^{\dag}_{n-1}A^{-2}_n),
\end{gathered}
\end{equation}
\noindent where the ``dot'' denotes derivative with respect to time. Similar expressions, compatible to ones above, arise from the set \eqref{eq:bti} and \eqref{eq:BTb}. 
Detailed discussion on the behaviour of these equations will be presented in a forthcoming work.

\section{Discussion}

\noindent The main aim now is to compute the time evolution of local operators using the time evolution operator $ \e{-itH} $, where $H$ 
in our case would be the Hamiltonian of the $q$-harmonic oscillator derived previously. It is clear that for any integrable system a more general 
description can be considered regarding the ``universal'' time evolution including all the time flows of the integrable hierarchy; in this case the 
object under consideration is $\e{-i {\mathrm T} {\mathfrak t}(\lambda)}$, where ${\mathfrak t}$ is the generating function of all integrals 
of motion and ${\mathrm T}$ the universal time.

The object under interest in this context would be the expectation value of local operators ${\cal O}_j \in \{b_j,\ b_j^{\dag}\}$:
\begin{equation}
{\cal E}(t)= \langle Q_f| {\cal O}_{j}(t) | Q_i\rangle, ~~~~~\mbox{where} ~~~~{\cal O}_j(t) = \e{-it H}\ {\cal O}_j\ \e{it H}.
\end{equation}
Expansion over the complete set of the energy eigenstate (Bethe state) then gives:
\begin{eqnarray}
{\cal E}(t) &=& \sum_{n, m} \langle Q_f|\Psi_n\rangle \e{-i(E_n-E_m)t}\langle \Psi_n | {\cal O}_{j} |\Psi_m \rangle  \langle \Psi_m | Q_i\rangle \nonumber \\
&=& \sum_{n, m} \e{-i(E_n-E_m)t}\Psi_n(\bar Q_f) {\cal O}_{nm} \bar \Psi_m(Q_i).
\end{eqnarray}
The use  of coherent states, which is briefly discussed in the subsequent section, leads to a semi-classical description of the time evolution problem. 
This issue however will be discussed in more detail in a forthcoming work.

The Bethe ansatz formulation is used for the derivation of the energy eigenvalues and eigenstates. In fact, the algebraic Bethe ansatz can be applied 
given that highest weight states exist, indeed locally one observes the existence of such states (recall also that $qv^2 +a^{\dag}a = q^{-1}v^2+ aa^{\dag} = 1$):
\begin{equation}
a_j |0\rangle_j =0, ~~~~v_j |0\rangle_j = q^{-\frac{1}{2}} |0\rangle_j.
\end{equation}
Then the global reference state is
\begin{equation}
|\Omega \rangle = \bigotimes_{j=1}^{N}|0\rangle_j.
\end{equation}

The monodromy matrix and the generic Bethe state are expressed as
\begin{equation}
	T(\lambda) = \lb \begin{matrix}
	{\mathrm A}	& {\mathrm B}	 \\
	{\mathrm C}	 & {\mathrm D}	
	\end{matrix} \rb, ~~~~~|\Psi_M(\{\lambda_k\})\rangle = \prod_{k=1}^M{\mathrm B}(\lambda_k)|\Omega \rangle. \nonumber
\end{equation}
The Bethe roots satisfy the Bethe ansatz equations (BAE). The BAEs are obtained as analyticity conditions imposed on the spectrum, and read as:
\begin{equation}
(-q^{-1})^N \e{2\lambda_i N} = \prod_{i\neq j} \frac{\sinh {\lambda_i -\lambda_j +i\mu}}{\sinh{\lambda_i -\lambda_j - i\mu}}.
\end{equation}
The algebraic Bethe ansatz method is used for the derivations of  the spectrum and BAE for the model under consideration.
 The spectrum of the transfer matrix in the periodic case reads as:
\begin{equation}
\Lambda(\lambda) = a_+^{N}(\lambda)\prod_{k=1}^M \frac{\sinh{\lambda - \lambda_k-i\mu}}{\sinh{\lambda - \lambda_k}} +
 (-1)^Na_-^N(\lambda)\prod_{k=1}^M \frac{\sinh{\lambda - \lambda_k+i\mu}}{\sinh{\lambda - \lambda_k}}, \label{sp1}
\end{equation}
where $a_{\pm}(\lambda) = \e{\pm \lambda}$.

Similarly, in the case of the open model with diagonal boundary conditions (we have considered here for simplicity both $K^{\pm} \propto {\mathbb I}$) t
he algebraic Bethe ansatz applies for the modified monodromy matrix and the generic Bethe states are expressed as
\begin{equation}
	{\cal T}(\lambda) = \lb \begin{matrix}
	{\mathfrak A}	& {\mathfrak B}	 \\
	{\mathfrak C}	 & {\mathfrak D}	
	\end{matrix} \rb, ~~~~~|\Psi_M(\{ \lambda_k \})\rangle = \prod_{k=1}^M {\mathfrak B}(\lambda_k)\ |\Omega\rangle.
\end{equation}

Use of the algebraic Bethe ansatz for the open model leads to the spectrum:
\begin{equation}
\begin{aligned}
	\Lambda(\lambda) &= q^N a_+^{2N}(\lambda)\prod_{k=1}^M \frac{\sinh{\lambda - \lambda_k-i\mu}}{\sinh{\lambda - \lambda_k}}\ 
\frac{\sinh{\lambda + \lambda_k}}{\sinh{\lambda - \lambda_k}} \\ 	&\qquad+q^{-N}a_-^{2N}(\lambda)\prod_{k=1}^M 
\frac{\sinh{\lambda - \lambda_k+i\mu}}{\sinh{\lambda - \lambda_k}}\ \frac{\sinh{\lambda + \lambda_k+2i\mu}}{\sinh{\lambda - \lambda_k}},
\end{aligned} \label{eq:sp2}
\end{equation}
\noindent and the corresponding BAEs read as
\begin{equation}
 \e{4\lambda_i N} = \prod_{i\neq j} \frac{\sinh {\lambda_i -\lambda_j +i\mu}}{\sinh{\lambda_i -\lambda_j - i\mu} }\  
\frac{\sinh {\lambda_i +\lambda_j +i\mu}}{\sinh{\lambda_i+\lambda_j - i\mu} }.
\end{equation}
\noindent Having at our disposal the spectrum and the corresponding Bethe ansatz equations for both periodic and open spin 
chains we can proceed with the computation of time expectation values. It is clear that the study of the Bethe ansatz equations in 
the thermodynamic limit will be most relevant in this setting (see e.g. \cite{caux}).

 An efficient way to deal with the time evolution of a quantum system is the use of coherent states. These have been extensively used in 
the context of integrable models, with significant applications for instance in condensed matter and string theory. Here we shall use the
 $q$-coherent states associated also to $q$-Hermite polynomials (see e.g. \cite{q-coherent} and references therein). 
The quantum algebra (\ref{eqs:AL_Comm}) can be re-expressed as follows, after a suitable rescaling of the $b,\ b^{\dag}$ operators:
\begin{equation}
	b\ b^{\dag} - q^{2} b^{\dag}\ b = 1.
\end{equation}
\noindent The local vacuum and the general eigenstate of the local operator $b^{\dag}\ b$ are then given by
\begin{equation}
\begin{aligned}
	b\ |0\rangle = 0, \qquad&\qquad b^{\dag}\ |0\rangle = |1 \rangle, \\
	|n \rangle = \frac{b^{\dag n}}{\sqrt{[n]!}}\ |0 \rangle, \qquad&\qquad \langle n| = \langle 0|\ \frac{b^n}{\sqrt{[n]!}},
\end{aligned}
\end{equation}
\noindent where we define the $q$-factorial $[n]!$ in terms of $[n] = \frac{q^{2n} - 1}{q^2 - 1}$ and $[n]! = \prod_{j=1}^n\ [j]$. 
The coherent state is then defined as:
\begin{equation}
	|z\rangle = \sum_n \frac{(z b^{\dag})^n}{[n]!}\ |0 \rangle = \exp_q(z b^{\dag})\ |0 \rangle, ~~~~~~
\langle z| = \langle 0|\ \sum_n \frac{(z b)^n}{[n]!} =  \langle 0| \exp_q(z b).
\end{equation}
\noindent These states have the advantage of providing a natural semi-classical description of the system under 
study as will become clear below. Indeed, coherent states are endowed with the following practical properties
\begin{equation}
	b\ |z \rangle = z\ |z\rangle, ~~~~~\langle z|\ b^{\dag} = \langle z|\ z^*,
\end{equation}
\noindent and:
\begin{equation}
	\langle z | z' \rangle = \exp_q(z^* z'), ~~~~\int \frac{dz dz^*}{2\pi i}\ W(|z|^2)\ |z \rangle\langle z| = {\mathbb I}.
\end{equation}
\noindent The weight $W$ is derived in terms of $\exp_q$ functions (see \cite{q-coherent}).

We are dealing here with an $N$-body quantum mechanical system, therefore we shall need a ``global'' space coherent state 
$|Z\rangle = \bigotimes_{j=1}^N\ |z_j\rangle$, then the resolution of the unit becomes:
\begin{equation}
\begin{gathered}
	\int {{\cal D} Z\ \cal D}Z^*\ {\cal W}(|Z|^2)\ |Z \rangle\langle Z| = {\mathbb I}, \\
	{\cal W}(|Z|^2) = \prod_{j=1}^N W(|z_j|^2), ~~~~~{\cal D}Z\ {\cal D}Z^* = \prod_{j=1}^N \frac{dz_j dz_j^*}{2\pi i} .
\end{gathered}
\end{equation}

\noindent Indeed, let us now consider the object of interest
\begin{equation}
	{\cal G} = \langle \Psi_f| \e{-it H} |\Psi_i\rangle.
\end{equation}
\noindent Inserting the complete set of coherent states we then obtain in the typical path integral formulation
\begin{equation}
\begin{gathered}
	{\cal G} = \int{\cal D {\mathbb Z}}\ {\cal D {\mathbb Z}^*}\ \Psi_f^*(Z_f) \Psi_i(Z_i)\ {\cal W}({|\mathbb Z}|^2)\
prod_{j=1}^N \prod_{\alpha=0}^{M+1}\exp_q(z^*_{j a+1}z_{ja})\  \e{-i\delta \sum_{\alpha}\langle H_{\alpha} \rangle}, \\
	{\cal D}{\mathbb Z}\ {\cal D}{\mathbb Z}^* = \prod_{j=1}^N \prod_{\alpha=0}^{M+1} \frac{ dz_{ja} dz^*_{ja}}{2\pi i}, 
~~~~~{\cal W}({|\mathbb Z}|^2) = \prod_{j=1}^N \prod_{\alpha=0}^{M+1} W(|z_{ja}|^2),
\end{gathered} \label{eq:g1}
\end{equation}
\noindent with the time boundary conditions $Z_{M+1} =  Z_f,\ Z_0 = Z_i$, where now the index $\alpha$ is a discrete time index.
 In our case here the associated Hamiltonian is given as $ qH^+ +q^{-1} H^{-} $ and hence,
\begin{equation}
	\langle H_{\alpha} \rangle = q \sum_{j=1}^N z^*_{j+1 \alpha+1}\ z_{j\alpha} + q^{-1} \sum_{j=1}^N z^*_{j \alpha+1}\ z_{j+1 \alpha}.
\end{equation}

\noindent The isotropic analogue of the latter expression (e.g. for the discrete NLS model) becomes, given that $\exp_q \to \exp$, and $W(|z|^2) = \exp(-z z^*)$,
 and after considering the continuum time limit:
\begin{equation}
	{\cal G} = \int{\cal D {\mathbb Z}}\ {\cal D {\mathbb Z}^*}\ \Psi_f^*(Z_f) \Psi_i(Z_i)\ \e{i \int_{t_i}^{t_f} dt \sum_j 
\Big (-\frac{i}{2} {\partial_t} z^*_{j} z_{j} + \frac{i}{2} {\partial_t} z_{j }z^*_{j}- \langle h^{(n)}_j \rangle\Big )}\ \e{-\frac{1}{2} \sum_j (|z_j(t_f)|^2)}, \label{eq:g2}
\end{equation}
\noindent where in general $H^{(n)} = \sum_j h_j^{(n)}$ is one of the conserved quantities of the hierarchy associated to the time flow $t_n$.

The latter computation of course can be generalized for the ``universal" time flow, where in the expression above $H^{(n)} \to {\mathfrak t}(\lambda)$.
 In the case of imaginary time the latter provides the partition function of the 2D statistical system
\begin{equation}
	{\cal Z} = \tr{\e{-\beta H}} = \int {\cal D} \zeta\ \langle \zeta| \e{-\beta H}|\zeta \rangle.
\end{equation}
\noindent The next natural step is explicit computations via the appropriate differential or difference operator, whose determinant will be used 
for the computation of the partition function of the system under study. Detailed derivations on discrete and continuum NLS model, associated 
to all time flows, in particular in the presence of time-like and space-like defects and boundaries will be presented elsewhere.
For some recent findings on  explicit computations of the path integral for many body quantum mechanical systems by 
means of stochastic analysis arguments we refer the interested reader to \cite{stoch}.

We have considered the quantisation of the auxiliary linear problem and the associated Darboux-B\"acklund transformation. In this setting we derived
 the quantum hierarchy of the time components of the Lax pairs in the case of both periodic and open integrable boundary conditions.
 Moreover, having identified via our generic construction the quantum Lax pair for the $q$-oscillator model, we were able to derive the quantum
 Darboux transformation and hence the quantum BT. We worked out explicitly both the time independent and the time dependant part of the BT. 
The time part of the BT provides further information regarding the time evolution of the degrees of freedom of the corresponding Darboux matrix,
 and in fact by simply considering the $t$ part of the BT we recover the information provided by the time independent part as well. We should 
emphasize once more that our description is a typical Heisenberg time evolution picture, however possible links to matrix models (random matrices)
\cite{matrix1, matrix, random} via the discrete space time expression \eqref{eq:g1} can be explored. In any case, keeping the 
`time slicing'' picture reflected in expression \eqref{eq:g1} direct analogies to discrete time integrable models at the level of the 
completely discrete Lax pair can be made. Indeed, recall the fully discrete auxiliary linear problem described as
\begin{equation}
\begin{aligned}
	\Psi(\alpha, n + 1) &= L(\alpha, n)\ \Psi(\alpha, n), \\
	\Psi(\alpha + 1, n) &= {\mathbb A}(\alpha, n)\ \Psi(\alpha, n),
\end{aligned}
\end{equation}
\noindent where $\alpha$ is the discrete time index and $n$ the discrete space index, and in the continuum limit $a\to t,\ n \to x$. 
In the usual setting of vertex models (2D lattices) $L={\mathbb A}$, and the global coherent state will be expressed as 
$|{\mathbb Z} \rangle = \bigotimes_{j=1}^N \bigotimes_{\alpha=1}^M |z_{an} \rangle$, which shows direct analogy to 
\eqref{eq:g1}, where both the space and time discretisations are kept. In the continuum time limit of course one recovers 
the semi-discrete case considered here.

The pertinent question is the interpretation of the quantum Darboux-B\"acklund transformation. In \cite{sklyanin, korff} the 
quantum BT is seen as a quantum canonical transformation and the $Q$-operator is indeed the generating function of 
this transformation. However, it is also known that quantum canonical transformations can be treated by means of suitable 
squeezed states (see e.g. \cite{squeez} and references therein), and this is an interesting direction to pursue within the 
present frame. Let us also recall the classical picture associated to the BT, which is rather closer to our perspective and is 
also most relevant to the context of super-symmetric quantum mechanics (see e.g. \cite{sasaki} and references therein). 
Indeed, at the classical level the BT can be seen as a canonical transformation that relates two distinct solutions of the same 
PDE (or different PDEs, hetero-BT). Let us now ask the same question at the level of Hamiltonian evolution.  Let ${\mathbb D}$ 
be the Darboux matrix that relates two Hamiltonians with two different potentials; in the integrable PDEs frame these potentials 
can be two distinct solutions of the same PDE. We focus on the time evolution of the two distinct Hamiltonians and consequently 
the Darboux transformation
\begin{equation}
\begin{gathered}
	i \partial_t \Psi =H\ \Psi, ~~~~i \partial_t \tilde \Psi =\tilde H\ \tilde \Psi, \\
	H =-\partial_x^2+V(x), ~~~~~\tilde H=-\partial_x^2+\tilde V(x), ~~~~\tilde \Psi = {\mathbb D}\ \Psi.
\end{gathered}
\end{equation}

\noindent The equations above lead to the time evolution equation for ${\mathbb D}$:
\begin{equation}
	i\partial_t {\mathbb D} = \tilde H\ {\mathbb D} - H\ {\mathbb D}. \label{eq:tt}
\end{equation}
\noindent The significant issue for us is the understanding of the Darboux-matrix as described above for $N$-body Hamiltonians. 
In  general, even in the case of the one particle Hamiltonian, the  transformation ${\mathbb D}$ can be a differential or an integral 
operator whose form can be identified after solving \eqref{eq:tt} for known $H$ and $\tilde H$.

In general, the path integral quantisation scheme in the context of $N$-body integrable models can be utilised to provide significant 
connections with results already obtained for instance via the Bethe ansatz formulation, or facilitate certain computations regarding 
for example the derivation of expectation values. Immediate links with conformal field theories, diffusion reaction models \cite{cardy, stoch}
as well matrix models and random matrices (see e.g \cite{matrix1, matrix, random}) can  also be further pursued in this context, in 
particular in the presence of non-trivial boundary conditions. Finally, a natural question to address is the identification of the quantum 
hetero-BT in the quantum Liouville theory \cite{qLiouville}. Quantization of the classical Darboux hetero-BT between the Liouville theory 
and the free massless theory found in \cite{doikou-findlay} is a work in progress. We hope to address the aforementioned significant
matters soon in forthcoming investigations.

\vskip 0.2cm

\noindent {\bf Acknowledgements}: We are indebted to C. Korff for valuable discussions on quantum B\"acklund 
transformations and the $Q$-operator, and also for sharing his previous results on the matter \cite{korff}. 
We would like to thank P. Adamopoulou, D. Palazzo, and R. Weston for illuminating comments and insights. 
We are also grateful to J. Avan for crucial comments and suggestions regarding the manuscript. A.D. 
would like to thank University of Cergy-Pontoise for kind hospitality. I.F. thanks EPSRC for financial support via a DTA studentship.

\appendix

\section{The ${\mathbb B}$-operators: closed spin chain}

\noindent Knowing the Hamiltonian, we only need to derive the corresponding $ {\mathbb B}_n $ matrix, which will similarly be constructed as:
\begin{equation}
	{\mathbb B}_n = ({\mathbb B}^{(+, 0)}_n)^{-1} {\mathbb B}^{(+, 2)}_n + ({\mathbb B}^{(-, 0)}_n)^{-1} {\mathbb B}^{(-, 2)}_n. \nonumber
\end{equation}

\noindent After an appropriate rescaling of the $ R $-matrix, we can calculate these $ {\mathbb B}^{(\pm, k)}_n $ matrices. 
Looking first at the results from the $ u \to \infty $ limit, we get that:
\begin{align}
	& {\mathbb B}^{(+, 0)}_n = u^{-1} v_N ... v_1 \lb \begin{matrix}
		q & 0 \\
		0 & 1
	\end{matrix} \rb, \qquad\qquad {\mathbb B}^{(+, 1)}_n = 0, \nonumber \\
	& {\mathbb B}^{(+, 2)}_n = v_N ... v_1  \nonumber\\ &\times \lb\begin{matrix}
		u^{-1} (q\sum_{j \neq n - 1}^{N} b^{\dag}_{j + 1} b_j + b^{\dag}_n b_{n - 1}) - uq^{-1} & (q - q^{-1}) b^{\dag}_n \\
		(q - q^{-1}) b_{n - 1} & u^{-1} (\sum_{j \neq n - 1}^{N} b^{\dag}_{j + 1} b_j + qb^{\dag}_n b_{n - 1}) - u
	\end{matrix} \rb. \nonumber
\end{align}

\noindent The factor we are actually interested in calculating is $ ({\mathbb B}^{(+, 0)}_n)^{-1} {\mathbb B}^{(+, 2)}_n $, which is:
\begin{equation}
	{\mathbb B}^{+}_n = \lb \begin{matrix}
		\sum_{j \neq n - 1}^{N} b^{\dag}_{j + 1} b_j + q^{-1} b^{\dag}_n b_{n - 1} - u^2 q^{-2} & u(1 - q^{-2}) b^{\dag}_n \\
		u(q - q^{-1}) b_{n - 1} & \sum_{j \neq n - 1}^{N} b^{\dag}_{j + 1} b_j + qb^{\dag}_n b_{n - 1} - u^2
	\end{matrix} \rb. \nonumber
\end{equation}

Next, we need to calculate the ${\mathbb B}^{(-, k)}_n $ matrices, found by looking in the limit as $ u \to 0 $. These are:
\begin{align}
	&{\mathbb B}^{(-, 0)}_n = u v_N ... v_1 \lb \begin{matrix}
		1 & 0 \\
		0 & q^{-1}
	\end{matrix} \rb, \qquad\qquad {\mathbb B}^{(-, 1)}_n = 0, \nonumber \\
	&{\mathbb B}^{(-, 2)}_n = v_N ... v_1 \nonumber\\ & \times \lb \begin{matrix}
		u (\sum_{j \neq n - 1}^{N} b_{j + 1} b^{\dag}_j + q^{-1} b_n b^{\dag}_{n - 1}) - u^{-1} & (q - q^{-1}) b^{\dag}_{n - 1} \\
		(q - q^{-1}) b_n & u (q^{-1} \sum_{j \neq n - 1}^{N} b_{j + 1} b^{\dag}_j + b_n b^{\dag}_{n - 1}) - u^{-1} q
	\end{matrix} \rb. \nonumber
\end{align}

\noindent Looking at the combination $ ({\mathbb B}^{(-, 0)}_n)^{-1} {\mathbb B}^{(-, 2)}_n $, we get that:
\begin{equation}
	{\mathbb B}^{-}_n = \lb \begin{matrix}
		\sum_{j \neq n - 1}^{N} b_{j + 1} b^{\dag}_j + q^{-1} b_n b^{\dag}_{n - 1} - u^{-2} & u^{-1} (q - q^{-1}) b^{\dag}_{n - 1} \\
		u^{-1} (q^2 - 1) b_n & \sum_{j \neq n - 1}^{N} b_{j + 1} b^{\dag}_j + qb_n b^{\dag}_{n - 1} - u^{-2} q^2
	\end{matrix} \rb. \nonumber
\end{equation}

\section{The Hamiltonians $\&$ ${\mathbb B}$-operators: open spin chain}

\noindent The Hamiltonians can be found by expanding the generator $ \mathfrak{t} $ about powers of $ u $. Again, we can look 
at the two cases where $ \lambda \to \pm \infty $. First, doing so for the $ u \to \infty $ limit, the three lowest order terms are:
\begin{align}
	H^{(+, 0)} &= q^N v_N^2 ... v_1^2, \qquad\qquad H^{(+, 1)} = 0, \nonumber \\
	H^{(+, 2)} &= q^N v_N^2 ... v_1^2 \big( \sum_{n = 1}^{N - 1} (b^{\dag}_{n + 1} b_n + q^{-2} b_{n + 1} b^{\dag}_n) + 
b^{\dag}_1 b_1 + q^{-2} b_N b^{\dag}_N \big), \nonumber
\end{align}

\noindent while the three lowest order terms from the $ u \to 0 $ limit are:
\begin{align}
	H^{(-, 0)} &= q^{-N} v_N^2 ... v_1^2, \qquad\qquad H^{(-, 1)} = 0, \nonumber \\
	H^{(-, 2)} &= q^{-N} v_N^2 ... v_1^2 \left( \sum_{n = 1}^{N - 1} (b_{n + 1} b^{\dag}_n + q^2 b^{\dag}_{n + 1} b_n) +
 b_1 b^{\dag}_1 + q^2 b^{\dag}_N b_N \right). \nonumber
\end{align}

\noindent Considering the combination $ H^{\pm} = (H^{(\pm, 0)})^{-1} H^{(\pm, 2)} $, we get the physical 
Hamiltonians for each of the two limits:
\begin{align}
	H^{+} &= \sum_{n = 1}^{N - 1} (b^{\dag}_{n + 1} b_n + q^{-2} b_{n + 1} b^{\dag}_n) + (b^{\dag}_1 b_1 +
 q^{-2} b_N b^{\dag}_N), \nonumber \\
	H^{-} &= \sum_{n = 1}^{N - 1} (b_{n + 1} b^{\dag}_n + q^2 b^{\dag}_{n + 1} b_n) + (b_1 b^{\dag}_1 + 
q^2 b^{\dag}_N b_N). \nonumber
\end{align}

In the limit as $ u \to \infty $, we can find the first few matrices in the expansion of this generator:
\begin{align}
	{\mathbb B}^{(+, 0)}_1 &= q^N v_N^2 ... v_1^2 \lb \begin{matrix}
		q^2 & 0 \\
		0 & 1
	\end{matrix} \rb, \qquad\qquad {\mathbb B}^{(+, 1)}_1 = 0, \nonumber \\
	{\mathbb B}^{(+, 2)}_1 &= q^N v_N^2 ... v_1^2 \nonumber\\ & \times \lb \begin{matrix}
		b^{\dag}_1 b_1 + b_N b^{\dag}_N - u^2 - u^{-2} & (u + u^{-1})(q^2 - 1) b^{\dag}_1 \\
		(u + u^{-1})(1 - q^{-2}) b_1 & q^2 b^{\dag}_1 b_1 + q^{-2} b_N b^{\dag}_N + (q - q^{-1})^2 - u^2 - u^{-2}
	\end{matrix} \rb \nonumber \\
	&+ q^N v_N^2 ... v_1^2 \sum_{j = 1}^{N - 1} (b^{\dag}_{j + 1} b_j + q^{-2} b_{j + 1} b^{\dag}_j) \lb \begin{matrix}
		q^2 & 0 \\
		0 & 1
	\end{matrix} \rb. \nonumber
\end{align}

\noindent From these, we are primarily interested in the combination $ {\mathbb B}^{+}_1 = ({\mathbb B}^{(+, 0)}_1)^{-1} {\mathbb B}^{(+, 2)}_1 $, which is:
\begin{align}
	{\mathbb B}^{+}_1 &= -(u^2 + u^{-2}) \lb \begin{matrix}
		q^{-2} & 0 \\
		0 & 1
	\end{matrix} \rb + \lb \begin{matrix}
		q^{-2} b^{\dag}_1 b_1 & (u + u^{-1})(1 - q^{-2}) b^{\dag}_1 \\
		(u + u^{-1})(1 - q^{-2}) b_1 & q^2 b^{\dag}_1 b_1 + (q - q^{-1})^2
	\end{matrix} \rb \nonumber \\
	&\qquad+ \lb \sum_{j = 1}^{N - 1} (b^{\dag}_{j + 1} b_j + q^{-2} b_{j + 1} b^{\dag}_j) + q^{-2} b_N b^{\dag}_N \rb \lb \begin{matrix}
		1 & 0 \\
		0 & 1
	\end{matrix} \rb. \nonumber
\end{align}

\end{document}